\documentclass[conference]{IEEEtran}
\pdfoutput=1
\usepackage[pdftex]{graphicx}
\DeclareGraphicsExtensions{.pdf,.jpeg,.png}
\usepackage[cmex10]{amsmath}
\usepackage{amsfonts, amssymb}
\usepackage{cite} 
\usepackage{hyperref}
\usepackage{multirow}
\newcommand{\wt}[2]{W_{#1}[#2]}
\newcommand{\wh}[2]{\hat{W}_{#1}[#2]}
\newcommand{\ws}[3]{\textbf{W}_{#1;#2}^{#3}} 
\newcommand{\wsh}[3]{\hat{\textbf{W}}_{#1;#2}^{#3}}
\newcommand{\Nob}[3]{N_{ob,A_{#1}}^{(#2,#3)}}
\newcommand{\NobE}[3]{\overline{N}_{ob,A_{#1}}^{(#2,#3)}}

\newcommand{\Pob}[3]{P_{ob,A_{#1}}^{(#2,#3)}}
\newcommand{\pr}{\text{Pr}} 
\newcommand{\xstar}{n}
\newcommand{\ystar}{q}

\newcommand{\xiN}[1]{\xi_{#1}} 
\newcommand{\vf}{\hspace{-0.6 mm}} 
\newcommand{\eqs}{\vspace*{-1.5 mm}}

\begin{document}




\title{Amplify-and-Forward Relaying in Two-Hop Diffusion-Based Molecular Communication Networks}
\author{ 
\IEEEauthorblockN{Arman Ahmadzadeh\IEEEauthorrefmark{1}, Adam Noel\IEEEauthorrefmark{2}, Andreas Burkovski\IEEEauthorrefmark{1}, and Robert Schober\IEEEauthorrefmark{1}} 
\IEEEauthorblockA{\IEEEauthorrefmark{1}University of Erlangen-Nuremberg, Germany } 
\IEEEauthorblockA{\IEEEauthorrefmark{2}University of British Columbia, Canada} 
} 
\maketitle 

\begin{abstract} 
This paper studies a three-node network in which an intermediate nano-transceiver, acting as a relay, is placed between a nano-transmitter and a nano-receiver to improve the range of diffusion-based molecular communication. Motivated by the relaying protocols used in traditional wireless communication systems, we study amplify-and-forward (AF) relaying with fixed and variable amplification factor for use in molecular communication systems. To this end, we derive a closed-form expression for the expected end-to-end error probability. Furthermore, we derive a closed-form expression for the optimal amplification factor at the relay node for minimization of an approximation of the expected error probability of the network. Our analytical and simulation results show the potential of AF relaying to improve the overall performance of nano-networks.
\end{abstract}
\section{Introduction}
Molecular communication (MC) is a biocompatible approach for enabling communication among so-called nano-machines by exchanging information via molecules. Integrating communication capabilities expands the potential functionality of individual nano-machines such that communities of them, so-called nano-networks, can execute collaborative and challenging tasks in a distributed manner. Sophisticated nano-networks are expected to have various biomedical, environmental, and industrial applications \cite{NakanoB}.    

Diffusion-based MC is a passive and energy-efficient approach for communication among nano-machines where the transportation of molecules from a nano-transmitter to a nano-receiver in a fluid environment relies on free diffusion only; no additional infrastructure is required. However, one of the main drawbacks of diffusion-based MC is its limited range of communication, since the propagation time increases and the number of received molecules decreases with increasing distance. This makes communication over larger distances challenging. 

\begin{figure}[!t] 
	\centering
	\includegraphics[scale = 1.5]{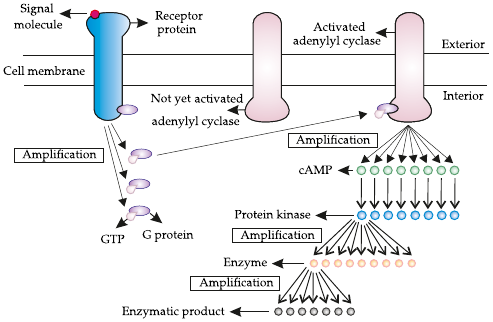}\vspace*{-2mm}
	\caption{Example for amplification of a signal at different stages of a typical cell-signaling process. Reproduced from \cite{Raven}.} 
	\label{Fig.cellSignaling}
	\vspace*{-4mm}
\end{figure}

One approach from conventional wireless communications that could be adapted for MC to aid communication with distant receivers is the use of relays. In fact, the relaying of information is already used by nature. In particular, cascades of signal amplification are a common way to relay information from the exterior of a cell to its interior \cite{Raven}. For example, the binding of a signal molecule to a cell surface receptor may activate a number of G protein molecules, each of which activates in turn a molecule of adenylyl cyclase, resulting in an excessive number of cAMP molecules. Subsequently, each cAMP molecule activates a protein kinase, which in turn generates several copies of a specific enzyme. Each enzyme can trigger a chemical reaction, resulting in a large amount of enzymatic product that can ultimately change the behavior of the cell, cf. Fig. \ref{Fig.cellSignaling}. For example, one rhodopsin pigment cell (the cells of the eye responsible for interpreting light and dark), when excited by a photon, can trigger the release of $10^{5}$ enzymatic products, i.e., cGMP molecules, so there is an amplification of five orders of magnitude.

\emph{Multihop relaying} among nano-machines has been studied in the existing MC literature; see \cite{Einolghozati1, Einolghozati2, Nakano2, Nakano3, Unluturk1, Balasubramaniam1, Walsh1, Bazargani1, Arman2, Arman1}. In \cite{Einolghozati1} and \cite{Einolghozati2}, a diffusion-based multihop network between bacteria colonies was analyzed, where each node of the network was formed by a population of bacteria. In \cite{Nakano2} and \cite{Nakano3}, the design and analysis of repeater cells in Calcium junction channels were investigated. In \cite{Unluturk1}, the rate-delay trade-off of a three-node nanonetwork was analyzed for a specific messenger molecule, polyethylene, and network coding at the relay node. The use of bacteria and virus particles as information carriers in a multihop network was proposed in \cite{Balasubramaniam1} and \cite{Walsh1}, respectively. Most recently, the authors in \cite{Bazargani1} investigated a two-hop network, where a relay node is installed to maximize the concentration of molecules at the destination. However, most prior works consider the transmission of a \emph{single} symbol (or bit) and, consequently, the effect of intersymbol interference (ISI), which is unavoidable if multiple symbols (bits) are transmitted, is not taken into account. An exception is our work in \cite{Arman2, Arman1}, where the performance of a decode-and-forward (DF) relaying protocol is studied. However, the complexity associated with full decoding at the relay may not be affordable in certain applications. On the other hand, installing a simple amplifier between a nano-transmitter and a nano-receiver may be easier to realize.         

In this paper, we assume that the transmitter nano-machine emits \emph{multiple} \emph{random} bits and investigate two different relaying protocols: variable-gain and fixed-gain amplify-and-forward (AF) relaying. In variable-gain AF relaying, the relay nano-machine amplifies the signal received from the nano-transmitter by a variable amplification factor, which may vary in each bit interval, to forward it to the nano-receiver. In fixed-gain AF relaying, the amplification factor at the relay nano-machine is constant for all bit intervals. The main contributions of this paper can be summarized as follows:
\begin{enumerate} 
 \item We derive closed-form expressions for the expected end-to-end error probability of variable-gain and fixed-gain AF relaying.
 \item We optimize the performance of variable-gain AF relaying by deriving a closed-form expression for the optimal amplification factor at the relay nano-machine. We propose a mechanism for variable-gain AF relaying where the relay nano-machine estimates the optimal amplification factor in each bit interval.
\item For fixed-gain AF relaying, the optimal amplification factor is derived by averaging the optimal amplification factor of variable-gain AF relaying over all bit intervals.  
\end{enumerate}    

The rest of this paper is organized as follows. In Section \ref{Sec.SysModandPre}, we introduce the system model and preliminaries of the error rate analysis. In Section \ref{Sec.Two-hopNetPerAna}, we derive the expected error probability of the two-hop network for both AF relaying protocols and the optimal amplification factor. Numerical results are presented in Section \ref{Sec. Simulations}, and conclusions are drawn in Section \ref{Sec.Con}. 
\section{SYSTEM MODEL AND PRELIMINARIES} 
\label{Sec.SysModandPre} 
In this section, we introduce the system model and some preliminaries required in the remainder of the paper. 
\subsection{System Model}
In this paper, we use the terms ``nano-machine'' and ``node'' interchangeably to refer to the devices in the network, as the term ``node" is commonly used in the relaying literature. We assume that a source ($S$) node and a destination ($D$) node are placed at locations $(0,0,0)$ and $(x_{D},0,0)$ of a 3-dimensional space, respectively. The relay ($R$) node is placed in the middle between node $S$ and node $D$ along the $x$-axis. We assume that node $D$ and node $R$ are spherical in shape with fixed volumes (and radii) $V_{D}$ $(r_{D})$ and $V_{R}$ $(r_{R})$, respectively, and that they are passive observers such that molecules can diffuse through them. For example, small, uncharged molecules, such as ethanol and urea, can enter and leave a cell by passive diffusion across the plasma membrane, see \cite{Raven}.

We assume that there are two distinct types of messenger molecules namely, type $A_{1}$ and type $A_{2}$, and that relay $R$ can detect type $A_{1}$ molecules, which are released by node $S$, and emits type $A_{2}$ molecules, which are detected by node $D$.
The number of molecules released of type $A_{f}, f \in \{ 1,2\}$, is denoted as $N_{A_{f}}$, and the concentration of type $A_{f}$ molecules at the point defined by vector $\vec{r}$ at time $t$ in molecule $\cdot$ $\text{m}^{-3}$ is denoted by $C_{A_{f}}( \vec{r}, t )$. We assume that the movements of individual molecules are independent.

The information that is sent from node $S$ to node $D$ is encoded into a binary sequence of length $L$, $\textbf{W}_{S} = \{ \wt{S}{1}, \wt{S}{2},..., \wt{S}{L} \}$. Here, $\wt{S}{j}$ is the bit transmitted by node $S$ in the $j$th bit interval with $\pr(\wt{S}{j}=1)=P_{1}$, and $\pr(\wt{S}{j}=0)=P_{0}=1-P_{1}$, where $\pr(\cdot)$ denotes probability. For compactness, we denote a subsequence transmitted by node $S$ by $\ws{S}{a}{b} = \{ \wt{S}{a},...,\wt{S}{b} \}$. The information bit detected at node $D$ in the $j$th bit interval is denoted by $\wh{D}{j}$. We adopt ON/OFF keying for modulation and a fixed bit interval duration of $T$ seconds. Node $S$ releases $N_{A_{1}}$ molecules at the beginning of the bit interval to convey information bit ``1'', and no molecules to convey information bit ``0''. Furthermore, we consider a full-duplex scheme, where reception and transmission occur simultaneously at the relay node, i.e., in each bit interval, relay $R$ receives a signal from node $S$, and releases an amplified version of the signal received in the previous bit interval to node $D$.

\subsection{Preliminaries} 
In the following, we consider a single communication link between a transmitting node $\xstar$ and a receiving node $\ystar$, and review the reception mechanism at node $\ystar$, cf. \cite{NoelJ2}. The two nodes can form the source-to-relay, ($\xstar = S, \ystar = R$), or relay-to-destination, ($\xstar = R, \ystar = D$), hop in our system model. For conciseness of presentation in this subsection, we drop the subscript $f$ and denote the type of molecule released by node $\xstar$ and detected at node $\ystar$ by $A$. 

The independent diffusion of molecules through the environment can be described by Fick's second law as
\begin{equation}
	\label{Eq. Fick's Second Law} 
	\frac{\partial C_{A}(\vec{r},t)}{\partial t} = D_{A} \nabla^2 C_{A}(\vec{r},t), \eqs
\end{equation} 
where $D_{A}$ is the diffusion coefficient of $A$ molecules in $\frac{\text{m}^2}{\text{s}}$. Assuming that node $\xstar$ is an impulsive point source, and emits $N_{A}$ molecules at the point defined by vector $\vec{r}_{\xstar}$ into an infinite environment at time $t' = 0$, the expected local concentration at the point defined by vector $\vec{r}$ and at time $t$ is given by \cite[Eq. (4)]{NoelJ1} 
\begin{equation}
	\label{Eq. Local Concentration} 
	C_{A}(\vec{r},t) = \frac{N_{A}}{(4 \pi D_{A} t)^{3/2}} \exp \left( -\frac{\vert \vec{r} - \vec{r}_{\xstar}\vert^2}{4D_{A}} \right). 
\end{equation} 

It is shown in \cite{NoelJ1} that the number of molecules observed within the volume of node $\ystar$, $V_{\ystar}$, at time $t$ due to one emission of $N_{A}$ molecules at $\vec{r}_{\xstar}$ at $t' = 0$, i.e., $\Nob{}{\xstar}{\ystar}(t)$, can be accurately approximated as a Poisson random variable (RV) with time-varying mean given by
\begin{equation}
	\label{Eq. NobE} 
	\NobE{}{\xstar}{\ystar}(t) = C_{A}(\vec{r}_{\ystar},t)V_{\ystar},
\end{equation}
where $\vec{r}_{\ystar}$ is the vector from the origin to the center of node $\ystar$. Eq. (\ref{Eq. NobE}) was derived under the uniform concentration assumption, i.e., it was assumed that node $\ystar$ is a point observer or that the concentration throughout its volume is uniform and equal to that at its center; see \cite{NoelPro1}. This is a valid assumption when node $\ystar$ is far from node $\xstar$. The probability of observing a given $A$ molecule, emitted by node $\xstar$ at $t'=0$, inside $V_{\ystar}$ at time $t$, i.e., $\Pob{}{\xstar}{\ystar}(t)$, is given by (\ref{Eq. NobE}) when setting $N_{A}=1$, i.e.,
\begin{equation}
	\label{Eq. Pob} 
	\Pob{}{\xstar}{\ystar}(t) = \frac{V_{\ystar}}{(4 \pi D_{A} t)^{3/2}} \exp \left( -\frac{\vert \vec{r}_{\ystar} -\vec{r}_{\xstar}\vert^2}{4D_{A}} \right). 
\end{equation}
 
For reception, we adopt the family of weighted sum detectors introduced in \cite{NoelJ2}, where the receiving node takes multiple samples within a single bit interval, and adds up the individual samples with a certain weight assigned to each sample. When detection is required at node $\ystar$, i.e., when $(\xstar = R, \ystar = D)$, then the sum is compared with a decision threshold. For simplicity, we assume equal weights for all samples. The decision in the $j$th bit interval is given by \cite[Eq. (37)]{NoelJ2} 
\begin{equation}
	\label{Eq.Reception} 
	\wh{\ystar}{j} = \begin{cases} 
	1 &\mbox{if } \sum_{m=1}^{M} \Nob{}{\xstar}{\ystar}(t(j,m)) \geq \xi_{\ystar}, \\
	0 &\mbox{otherwise,} 
			\end{cases}
\end{equation} 
where $\xi_{\ystar}$ is the detection threshold of node $\ystar$, and we assume that node $\ystar$ takes $M$ equally-spaced samples in each bit interval. The sampling time of the $m$th sample in the $j$th bit interval is $t(j,m) = (j-1)T_{B} + t_{m}$, where $t_{m}=mt_{0}$ and $t_{0}$ is the time between two successive samples. We assume that $t_{0}$ is large enough for samples to be independent. $\Nob{}{\xstar}{\ystar}(t(j,m))$ is a Poisson RV with mean $\NobE{}{\xstar}{\ystar}(t(j,m))$ for any individual sample. Thus, the sum of all samples in the $j$th bit interval, $\Nob{}{\xstar}{\ystar}[j]=\sum_{m=1}^{M} \Nob{}{\xstar}{\ystar}(t(j,m))$, is also a Poisson RV whose \emph{mean} is the sum of the \emph{means} of the individual samples, i.e., $\NobE{}{\xstar}{\ystar}[j] = \sum_{m=1}^{M}\NobE{}{\xstar}{\ystar}(t(j,m))$. Due to the independent movement of the molecules, node $\ystar$ observes molecules that were emitted at the start of the current or any prior bit interval. As a result, the number of molecules observed within $V_{\ystar}$ in the $j$th bit interval due to the transmission of sequence $\ws{\xstar}{1}{j}$, $\Nob{}{\xstar}{\ystar}[j]$, is also a Poisson RV with mean 
\begin{equation}
	\label{Eq. NobE Cumulative} 
	\NobE{}{\xstar}{\ystar}[j] = \sum_{i=1}^{j} N_{A}[i] \sum_{m=1}^{M} \Pob{}{\xstar}{\ystar}((j-i)T + t_{m}),
\end{equation} 
where $N_{A}[i]$ is the number of molecules released at node $\xstar$ at the beginning of the $i$th bit interval.

The cumulative distribution function (CDF) of the weighted sum in the $j$th bit interval is given by \cite[Eq. (38)]{NoelJ2} 
\begin{align}
	\label{Eq. CDF} 
	\pr \left( \Nob{}{\xstar}{\ystar}[j] < \xi_{\ystar} | \ws{\xstar}{1}{j} \right) &= \exp (-\NobE{}{\xstar}{\ystar}[j]) \nonumber \\
     & \times \sum_{\omega=0}^{\xi_{\ystar}-1} \frac{\left( \NobE{}{\xstar}{\ystar}[j] \right)^{\omega}}{\omega!}.
\end{align}
\section{PERFORMANCE ANALYSIS OF AF RELAYING}
\label{Sec.Two-hopNetPerAna} 
In this section, we evaluate the expected error probability of the two AF relaying protocols, i.e., variable-gain and fixed-gain AF relaying, respectively.

\subsection{Variable-Gain AF Relaying}
Node $S$ emits type $A_{1}$ molecules, which have diffusion coefficient $D_{A_{1}}$ and can be recognized by relay node $R$. In response to the received $A_{1}$ molecules, the relay emits type $A_{2}$ molecules having diffusion coefficient $D_{A_{2}}$. Node $S$ releases a fixed number of molecules, $N_{A_{1}}$, and no molecules to transmit bit ``1'' and bit ``0'' at the beginning of a bit interval, respectively. However, the number of molecules released by relay node $R$ at the beginning of a bit interval, $N_{A_{2}}[\cdot]$, varies in each bit interval. In particular, $N_{A_{2}}[\cdot]$ depends on the time-varying and random number of $A_{1}$ molecules observed at the relay node in the previous bit interval.   

Assuming full-duplex relaying at the relay node, the transmission of information bit $\wt{S}{j}$ from node $S$ to node $D$ has two phases. In the first phase, at the beginning of the $j$th bit interval, node $S$ transmits information bit $\wt{S}{j}$, and node $R$ releases $N_{A_{2}}[j]$ molecules concurrently to amplify and forward the message received in the previous bit interval, i.e., 
\begin{equation}
	\label{Eq. NumReleasedMolAtR}
	N_{A_{2}}[j] = k[j] \Nob{1}{S}{R}[j-1],
\end{equation} 
where $k[j]$ is the amplification factor of node $R$ in the $j$th bit interval.

At the end of the $j$th bit interval, node $D$ makes a decision on its received signal, and node $R$ receives signal $\Nob{1}{S}{R}[j]$. In the second phase, node $S$ transmits information bit $\wt{S}{j+1}$, and relay node $R$ emits $N_{A_{2}}[j+1] = k[j+1] \Nob{1}{S}{R}[j]$ molecules, which convey the information regarding $\wt{S}{j}$. At the end of the $(j+1)$th bit interval, node $D$ makes a decision on $\wt{S}{j}$, $\wh{D}{j+1}$. Thus, node $D$ receives the $j$th bit with a one bit interval delay. The total duration of transmission for a sequence of length $L$ is $(L+1)T$. 

\subsection{Expected Error Probability}
Given $\wt{S}{j}$, an error occurs in the $(j+1)$th bit interval if $\wh{D}{j+1} \neq \wt{S}{j}$. Thus, the error probability of the $j$th bit, $P_{e}[j|\wt{S}{j}]$ can be written as 
\begin{equation}
	\label{Eq. ErrorExpCurBit} 
	P_{e}[j|\wt{S}{j}] = \pr( \wt{S}{j} \neq \wh{D}{j+1}).	
\end{equation}
        
Let us assume that $\ws{S}{1}{j-1}$ is given. Then, the error probability of the $j$th bit when $\wt{S}{j}=1$ and $\wt{S}{j}=0$ can be written as 
\begin{align}
	\label{Eq. ErrorExpSeqW1} 
	  P_{e}[j|\wt{S}{j}  =  1, \ws{S}{1}{j-1}] =&\; \nonumber \\
	  &\; \hspace{-2.5 cm} \pr (\Nob{2}{R}{D}[j+1]  <  \xiN{D} | \wt{S}{j}  =  1, \ws{S}{1}{j-1}),
\end{align} 
and 
\begin{align}
	\label{Eq. ErrorExpSeqW0} 
	  P_{e}[j|\wt{S}{j}  =  0, \ws{S}{1}{j-1}] =&\; \nonumber \\
	  &\; \hspace{-2.5 cm} \pr (\Nob{2}{R}{D}[j+1]  \geq  \xiN{D} | \wt{S}{j}  =  0, \ws{S}{1}{j-1}),
\end{align} 
respectively. Hence, the expected error probability is given by 
\begin{align}
	\label{Eq. ErrorExpSeq} 
	  P_{e}[j|\ws{S}{1}{j-1}] =&\; P_{1}P_{e}[j|\wt{S}{j} \vf = \vf 1, \ws{S}{1}{j-1}] \nonumber \\
	  &\ + P_{0}P_{e}[j|\wt{S}{j} \vf = \vf 0, \ws{S}{1}{j-1}].
\end{align} 

In (\ref{Eq. ErrorExpSeqW1}), for given $\Nob{1}{S}{R}[i], i \in \{1, \cdots, j \}$, $\Nob{2}{R}{D}[j+1]$ is a Poisson RV with mean 
\begin{equation}
\label{Eq. PoissonMean} 
\NobE{2}{R}{D}[j+1] \vf = \vf \vf \sum_{i=1}^{j} k[i+1]\Nob{1}{S}{R}[i] \vf \sum_{m=1}^{M} \vf \Pob{2}{R}{D}((j-i)T + t_{m}),
\end{equation} 
where, in turn, $\Nob{1}{S}{R}[i]$ is also a Poisson RV. Thus, the conditional probability in (\ref{Eq. ErrorExpSeqW1}) can be evaluated as 
\begin{multline}
	\label{Eq. ConditionlProbEndToEnd} 
	P_{e}[j|\wt{S}{j}=  1, \ws{S}{1}{j-1}]= \\
	\sum_{\gamma_{1}=0}^{\infty} \cdots \sum_{\gamma_{j}=0}^{\infty} \pr \left( \Nob{2}{R}{D}[j+1] \vf < \vf \xiN{D} \big| \Nob{1}{S}{R}[1] = \gamma_{1}, \cdots, \right. \\
	\left. \Nob{1}{S}{R}[j] = \gamma_{j}, \wt{S}{j}=1, \ws{S}{1}{j-1} \right) \times \pr \left(  \Nob{1}{S}{R}[1] = \gamma_{1}, \right. \\
	 \left. \cdots, \Nob{1}{S}{R}[j] = \gamma_{j} \big| \wt{S}{j}=1, \ws{S}{1}{j-1} \right),
\end{multline} 
where $\gamma_{i}$, $i \in \{1, \cdots, j \}$, is one realization of Poisson RV $\Nob{1}{S}{R}[i]$ for given $\{ \wt{S}{j}=1, \ws{S}{1}{j-1} \}$. The first probability inside the sums in (\ref{Eq. ConditionlProbEndToEnd}) is the conditional CDF of Poisson RV $\Nob{2}{R}{D}[j+1]$ which can be evaluated based on (\ref{Eq. CDF}) given $\NobE{2}{R}{D}[j+1]$. This time-varying mean, in turn, can be evaluated based on (\ref{Eq. PoissonMean}) after substituting $\Nob{1}{S}{R}[i]$ with $k[i+1] \gamma_{i}$. The second probability inside the sums in (\ref{Eq. ConditionlProbEndToEnd}) is the conditional joint probability mass function (PMF) of Poisson RVs $\Nob{1}{S}{R}[i], i \in \{1, \cdots, j \}$, for given $\{ \wt{S}{j}=1, \ws{S}{1}{j-1} \}$, where due to the independent observation of molecules in each bit interval, we have
\begin{multline} 
	\label{Eq. PMF}
	\pr \left(  \Nob{1}{S}{R}[1] \vf = \vf \gamma_{1}, 
	\cdots, \Nob{1}{S}{R}[j] \vf = \vf \gamma_{j} \big| \wt{S}{j}=1, \ws{S}{1}{j-1} \right)  \\
	 = \prod_{i=1}^{j} \left( \frac{ \exp(- \NobE{1}{S}{R}[i]) (\NobE{1}{S}{R}[i])^{\gamma_{i}} }{\gamma_{i}!} \right),  
\end{multline}
where $\NobE{1}{S}{R}[i]$ can be evaluated based on (\ref{Eq. NobE Cumulative}) after substituting $N_{A}[i]$, $\xstar$, and $\ystar$ with $N_{A_{1}} \wt{S}{i}$, $S$, and $R$, respectively. Analogously, the conditional probability in (\ref{Eq. ErrorExpSeqW0}) can be evaluated by considering the complementary probability of (\ref{Eq. ConditionlProbEndToEnd}) after substituting $\wt{S}{j} = 1$ with $\wt{S}{j} = 0$. In (\ref{Eq. ConditionlProbEndToEnd}), there are infinitely many realizations for each Poisson RV $\Nob{1}{S}{R}[i]$. In order to keep the complexity of evaluation low, we consider only one random realization of each Poisson RV $\Nob{1}{S}{R}[i]$ with mean $\NobE{1}{S}{R}[i]$ for a given $\ws{S}{1}{i}$ for evaluation of (\ref{Eq. ConditionlProbEndToEnd}). Our simulation results in Section \ref{Sec. Simulations} confirm the accuracy of this approximation. The expected average error probability, $\bar{P}_{e}$, is obtained by averaging (\ref{Eq. ErrorExpSeq}) over all possible realizations of $\ws{S}{1}{j-1}$ and all bit intervals.      

\subsection{Optimal Amplification Factor}
In order to maximize the performance of variable-gain AF relaying, we derive a closed-form expression for the optimal amplification factor, $k_{opt}[\cdot]$, at the relay node for minimization of an approximation of the expected error probability of this network. Specifically, we approximate the random observation of molecules at node $R$ by the average number of observed molecules for evaluation of (\ref{Eq. ConditionlProbEndToEnd}), i.e., in (\ref{Eq. PoissonMean}), $\Nob{1}{S}{R}[i]$ is substituted with its mean $\NobE{1}{S}{R}[i]$. This approximation allows us to approximate $\Nob{2}{R}{D}[j+1]$ in (\ref{Eq. ErrorExpSeq}) by a Poisson RV with a deterministic mean given as         
\begin{align} 
\label{Eq. OveralExpectationOne}
	\NobE{2}{R}{D}[j \vf + \vf 1] \vf = \vf \vf \sum_{i=1}^{j} \vf k[i \vf + \vf 1] \NobE{1}{S}{R}[i] \vf \sum_{m=1}^{M} \vf \vf \Pob{2}{R}{D}((j\vf - \vf i)T  \vf +  \vf t_{m}). 
\end{align}
We now derive an expression for $k_{opt}[\cdot]$, for minimization of an approximation of (\ref{Eq. ErrorExpSeq}). To this end, by substituting $\NobE{1}{S}{R}[i]$ from (\ref{Eq. NobE Cumulative}), $\NobE{2}{R}{D}[j \vf + \vf 1]$ can be expressed as      
\begin{multline}
	\label{Eq. OveralExpectationTwo} 
	\NobE{2}{R}{D}[j \vf + \vf 1] \vf = \vf N_{A_{1}} \sum_{i=1}^{j} k[i \vf + \vf 1] \sum_{\omega =1 }^{i} \wt{S}{\omega} \\ \times \sum_{l=1}^{M} \vf \Pob{1}{S}{R}((i \vf - \vf \omega)T  \vf +  \vf t_{l})
	 \sum_{m=1}^{M} \vf \Pob{2}{R}{D}((j \vf - \vf i)T  \vf +  \vf t_{m}).
\end{multline} 
 
Eq. (\ref{Eq. OveralExpectationTwo}) can be rearranged to express $\NobE{2}{R}{D}[j+1]$ as a function of the most recent bit, $\wt{S}{j}$, and all prior transmitted bits $\left( \wt{S}{i}, i \in \{1, \cdots, j-1 \} \right)$, as 
\begin{multline}
	\label{Eq. Amplified-ISI} 
	\NobE{2}{R}{D}[j \vf + \vf 1] \vf = \vf N_{A_{1}} \sum_{i=1}^{j-1} k[i \vf + \vf 1] \sum_{\omega =1 }^{i} \wt{S}{\omega} \\  \times \sum_{l=1}^{M} \vf \Pob{1}{S}{R}((i \vf - \vf \omega)T  \vf +  \vf t_{l})  
	   \sum_{m=1}^{M} \vf \Pob{2}{R}{D}((j \vf - \vf i)T  \vf +  \vf t_{m})  +  k[j \vf + \vf 1] N_{A_{1}} \\ \times \sum_{\omega =1 }^{j-1} \wt{S}{\omega} 
 \sum_{l=1}^{M} \vf \Pob{1}{S}{R}((j \vf - \vf \omega)T  \vf +  \vf t_{l}) \sum_{m=1}^{M} \vf \Pob{2}{R}{D}(t_{m})  \\ 
	 \hspace{- 0.6 cm} + k[j+1] N_{A_{1}} \wt{S}{j} \sum_{l=1}^{M} \vf \Pob{1}{S}{R}(t_{l}) \sum_{m=1}^{M} \vf \Pob{2}{R}{D} (t_{m}),   
\end{multline} 
where the first term on the right hand side represents the ISI produced in the second hop, and the second term represents the expected number of molecules observed in $V_D$ due to the amplification of molecules originating from previous bit intervals and observed at node $R$ in bit interval $j$, $\wt{S}{i}, i < j$ (we refer to this sum as the \emph{amplified ISI of the first hop}). The third term is the expected number of molecules observed due to the transmission of the most recent bit $\wt{S}{j}$. Let us assume that the $k[i+1]$ are given. Considering (\ref{Eq. Amplified-ISI}) and the use of a fixed threshold at node $D$, $\xi_{D}$, we observe that decreasing $k[j+1]$ reduces the overall effect of ISI, and increases the probability of miss detection, i.e., $\pr(\wh{D}{j+1} \neq \wt{S}{j}| \wt{S}{j} =1 )$. On the other hand, increasing $k[j+1]$ enhances the ISI which in turn increases the probability of false alarm, i.e., $\pr(\wh{D}{j+1} \neq \wt{S}{j}| \wt{S}{j} =0 )$. Thus, the expected error probability in (\ref{Eq. ErrorExpSeq}) can be minimized by optimizing $k[j+1]$. The optimal $k[j+1]$ is provided in the following proposition.

\textit{Proposition 1:} Given $\ws{S}{1}{j-1}$, the optimal amplification factor at relay node $R$ at the beginning of the $(j+1)$th bit interval, $k_{opt}[j+1]$, which minimizes the approximate expected error probability of the $j$th bit, can be approximated as   
\begin{equation}
	\label{Eq. OptimalK} 
	k_{opt}[j+1] = \bigg\lfloor \frac{ \left( \xi_{D}-1 \right) \ln \left( \frac{B_{1}[j]}{B_{0}[j]} \times \sqrt[\xi_{D}-1]{ \frac{P_{1}B_{1}[j]}{P_{0}B_{0}[j]} }  \right) }{\sum_{l=1}^{M} \Pob{1}{S}{D}(t_{l}) \sum_{m=1}^{M} \Pob{2}{R}{D}(t_{m})} \bigg\rceil,
\end{equation} 
where $\lfloor \cdot \rceil$ is the nearest integer, $\ln(\cdot)$ is the natural logarithm, and $B_{b}[j+1], b \in \{0,1\}$ is obtained using (\ref{Eq. OveralExpectationTwo}) as 
\begin{multline}
	B_{b}[j] \big|_{ \wt{S}{j}=b, \ws{S}{1}{j-1}} = \sum_{\omega =1 }^{j-1} \wt{S}{\omega} \sum_{l=1}^{M} \vf \Pob{1}{S}{R}((j \vf - \vf \omega)T  \vf +  \vf t_{l}) \\ 
	 \hspace{-2 cm} \times \Pob{2}{R}{D} (t_{m}) + \wt{S}{j} \sum_{l=1}^{M} \vf \Pob{1}{S}{R}(t_{l}) \sum_{m=1}^{M} \vf \Pob{2}{R}{D} (t_{m}). 
\end{multline} 
\begin{IEEEproof}
$\Nob{2}{R}{D}[j+1]$ is approximated as a Poisson RV, and hence has a \emph{discrete} CDF which makes the optimization of $k[j+1]$ difficult. Fortunately, this CDF can be well approximated by a \emph{continous} regularized incomplete Gamma function as \cite[Eq. (1)]{Ilienko} 
\begin{equation}
	\pr(\Nob{2}{R}{D}[j \vf + \vf 1] < \xiN{D} | \ws{S}{1}{j}) \simeq \frac{\Gamma(\xiN{D},\NobE{2}{R}{D}[j \vf + \vf 1]| \ws{S}{1}{j} )}{\Gamma(\xiN{D})},
\end{equation} 
where $\Gamma(s, \rho)$ is the incomplete Gamma function given by $\Gamma(s,\rho) = \int_{\rho}^{\infty} e^{-t}t^{s-1} dt$ \cite[Eq. (6.5.3)]{abramowitz}. The Gamma function, $\Gamma(s)$, is a special case of the incomplete Gamma function with $\rho=0$. Using this approximation, we take the partial derivative of (\ref{Eq. ErrorExpSeq}) with respect to $k[j+1]$, solve the resulting equation for $k[j+1]$, and round the result to the nearest integer value $k_{opt}[j+1]$, which yields (\ref{Eq. OptimalK}). 
\end{IEEEproof} 


In the remainder of this subsection, we differentiate between two types of variable-gain AF relaying.

\textit{Type 1:} In this case, we assume that relay node $R$ adjusts its amplification factor in the $j$th bit interval to $k_{opt}[j]$. However, as shown in (\ref{Eq. OptimalK}), the evaluation of $k_{opt}[j]$ requires knowledge of the bits transmitted by node $S$. In order to cope with this uncertainty, relay node $R$ compares its received message in the $j$th bit interval, $\Nob{1}{S}{R}[j]$, with a decision threshold $\xiN{R}$ to estimate $\wt{S}{j}$. We denote this estimate by $\wh{R}{j}$, and for compactness the estimated information bits up to the current bit are denoted by $\wsh{R}{1}{j-1}$. 

Thus, given $\wsh{R}{1}{j-1}$, relay node $R$ estimates the optimal amplification factor in the $j$th bit interval, $\hat{k}_{opt}[j+1]$, via (\ref{Eq. OptimalK}) after substituting $\ws{S}{1}{j-1}$ with $\wsh{R}{1}{j-1}$, and releases $N_{A_{2}}[j+1]= \hat{k}_{opt}[j+1] \times  \Nob{1}{S}{R}[j]$, $A_{2}$ molecules at the beginning of the next $(j+1)$th bit interval. 

\textit{Remark 1:} The complexity associated with Type 1 variable-gain AF relaying is the same as that of DF relaying, since for estimation of $\wt{S}{j}$, i.e., $\wh{R}{j}$, relay node $R$ has to decode the received signal in each bit interval, which is not desirable.   

\textit{Type 2:} To reduce the computational complexity of Type 1 variable-gain AF relaying, we substitute $\hat{k}_{opt}[j+1]$ by $\bar{k}_{opt}[j+1]$, where $\bar{k}_{opt}[j+1]$ is obtained by averaging (\ref{Eq. OptimalK}) over all possible realizations of the sequence $\ws{S}{1}{j-1}$, i.e., 
\begin{equation}
	\label{Eq. OptimalKType2} 
	\bar{k}_{opt}[j+1] = \frac{1}{\mid \mathcal{W} \mid} \sum_{\mathcal{W}} k_{opt}[j+1], 
\end{equation}
where $\mathcal{W}$ is a set containing all possible realizations of $\ws{S}{1}{j-1}$, and $\mid\mathcal{W}\mid$ denotes the cardinality of set $\mathcal{W}$. Hence, for Type 2 variable-gain AF relaying the amplification factor in each bit interval can be computed offline. In Section \ref{Sec. Simulations}, we show that $\bar{k}_{opt}[\cdot]$ converges to an asymptotic value after several bit intervals.  

The overall end-to-end expected error probability of both Type 1 and Type 2 variable-gain AF relaying can be evaluated via (\ref{Eq. ErrorExpSeq}) after substituting $k[i+1], i \in \{1, \cdots, j\}$, with $\hat{k}_{opt}[i+1]$ and $\bar{k}_{opt}[i+1]$, respectively.      
\subsection{Fixed-Gain AF Relaying}
In fixed-gain AF relaying, the relay node $R$ adopts a bit-interval-independent, fixed amplification factor $k$ for forwarding the signal received from node $S$. Specifically, the amplification factor is denoted by $\bar{k}_{opt}$, where $\bar{k}_{opt}$ is obtained by averaging (\ref{Eq. OptimalK}) over all possible realizations of $\ws{S}{1}{j-1}$ \emph{and} over all bit intervals, i.e.,
\begin{equation}
	\label{Eq. OptimalKFixedGain} 
	\bar{k}_{opt} = \frac{1}{L \mid \mathcal{N} \mid} \sum_{j=1}^{L} \sum_{\mathcal{N}} k_{opt}[j+1], 
\end{equation} 
where set $\mathcal{N}$ contains all possible realizations of the sequence $\ws{S}{1}{L}$.

Fixed-gain AF relaying has a lower real-time computational complexity in comparison to both proposed variable-gain AF relay protocols, since no bit detection and no adjustment of the amplification factor ($k_{opt}[\cdot]$) are required at the relay node. Also, no prior knowledge of time-varying amplification factors in each bit interval ($\bar{k}_{opt}[\cdot]$) is required. The overall end-to-end expected error probability of fixed-gain AF relaying can be obtained from (\ref{Eq. ErrorExpSeq}) after substituting $k[i+1], i \in \{1, \cdots, j\}$, with $\bar{k}_{opt}$.                                 
\section{NUMERICAL RESULTS}
\label{Sec. Simulations} 
\begin{table}
	\begin{small}
		\begin{minipage}[b]{\linewidth} \hspace*{-1cm}
		\renewcommand{\arraystretch}{0.75} 
		\caption{SYSTEM PARAMETERS USED IN SIMULATIONS}
		\label{Table1}
		\centering
			\begin{tabular}{|c|c|c|} 
				\hline 
				\bfseries Parameter & \bfseries Symbol  & \bfseries Value \\ 
				\hline 
					Probability of binary 1 & $P_1$ & 0.5 \\ 
				\hline 
					Receiver distance & $x_{D}$ & 500 nm \\ 
				\hline 
					Length of transmitter sequence & $L$ & 50 \\ 
				\hline 
					Radius of relay node $R$ & $r_{R}$ & 45 nm \\ 
				\hline 
					Radius of node $D$ & $r_{D}$ & 45 nm \\ 
				\hline 
					Diffusion coefficient \cite{NoelJ1},\cite{NoelJ2} & $D_{A_{f}}$ & $4.365 \times 10^{-10} \frac{\text{m}^2}{\text{s}}$ \\
				\hline
			\end{tabular} 
		\end{minipage}\hspace*{1.2cm}
	\end{small}
	\vspace*{-4 mm}	
\end{table} 
In this section, we present simulation and analytical results for evaluation of the performance of the proposed relaying protocols. We adopted the particle-based stochastic simulator introduced in \cite{NoelJ1}. In our simulations, time is advanced in discrete steps of $t_{0}$, i.e., the time between two consecutive samples, where in each time step molecules undergo random motion. The environment parameters are listed in Table \ref{Table1}.
 
In order to focus on a comparison of the performance of the different relaying protocols, we keep the physical parameters of the relay and the receiver constant throughout this section. The parameters that we vary are the decision threshold $\xi_{D}$, the modulation bit interval $T$, the amplification factor $k$, and the frequency of sampling $1/t_{0}$.  

In the following, we refer to the case when no relay is deployed between node $S$ and node $D$ as the \textit{baseline case}. We adopt $N_{A_{1}} = 2500$ for the two-hop network. For a fair comparison between the two-hop case and the baseline case, we assume for the baseline case that for transmission of information bit ``1'', $N_{A_{1}} = 2500 + 2 \bar{N}_{A_{2}}$ molecules are released by $S$, where $\bar{N}_{A_{2}}$ is the average number of molecules released by node $R$ for transmission of one bit in the two-hop case.

\begin{figure}[!t]
	\centering
	\includegraphics[scale=0.27]{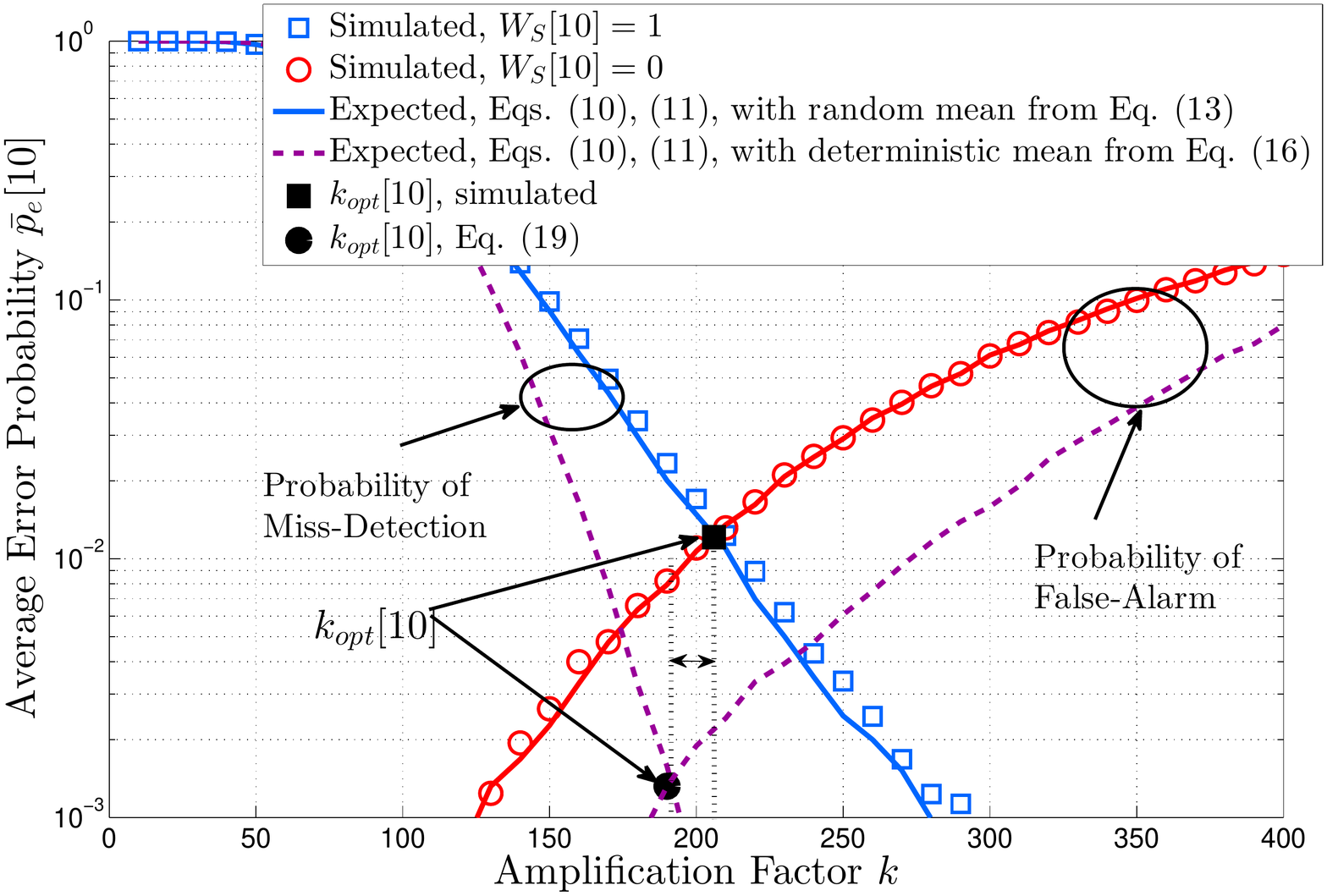}\vspace*{-6mm}
	\caption{Average error probability of 10th bit, given one randomly-chosen 9-bit sequence, vs. amplification factor $k$. The first 9 bits are ``101101001''.}
	\label{Fig.2}
	\vspace*{-6 mm}
\end{figure} 
In Fig. \ref{Fig.2}, the average error probability of the 10th bit, given one randomly-chosen 9-bit sequence, is found as a function of amplification factor $k$, for system parameters $M=10$, $T=400 \, \mu s$, and $\xiN{D}=20$. The simulations are averaged over $10^{5}$ independent transmissions of the chosen sequence. We see that increasing $k$ increases and decreases the probabilities of false-alarm and miss-detection, respectively, which confirms the existence of an optimal $k$ for the minimization of (\ref{Eq. ErrorExpSeq}). The expected error probabilities are evaluated using (\ref{Eq. ErrorExpSeqW1}) and (\ref{Eq. ErrorExpSeqW0}), when $\wt{S}{10}=1$ and $\wt{S}{10}=0$, respectively, and considering two approximations for $\NobE{1}{S}{R}[\cdot]$, i.e., Eqs. (\ref{Eq. PoissonMean}) and (\ref{Eq. OveralExpectationOne}). Although using the approximation leads to an underestimation of the expected error probability in Fig. \ref{Fig.2}, the difference between $k_{opt}[\cdot]$ obtained via (\ref{Eq. OptimalK}) and $k_{opt}[\cdot]$ found via simulation is small (less than 9\%).                 

\begin{figure}[!t]
	\centering
	\includegraphics[scale=0.27]{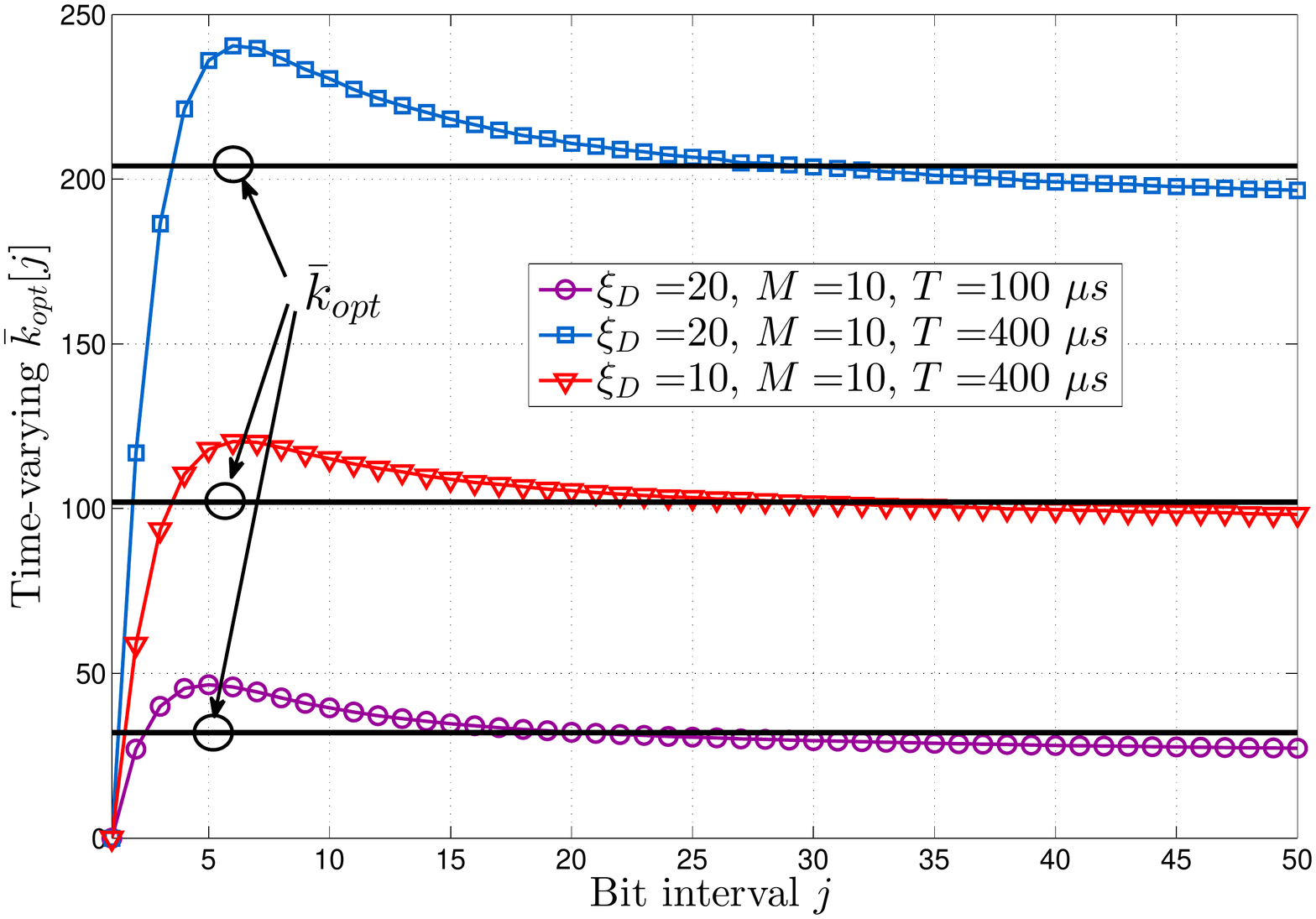}\vspace*{-6mm}
	\caption{Time-varying $\bar{k}_{opt}[j]$ as a function of bit interval.}
	\label{Fig.3}
	\vspace*{-7mm}
\end{figure}
In Fig. \ref{Fig.3} we show the amplification factor for Type 2 variable-gain AF relaying, $\bar{k}_{opt}[j]$, as a function of the bit interval $j$, for three different sets of system parameters. The results are averaged over $10^{5}$ independent realizations of $\ws{S}{1}{L}$. For the considered scenario, $\bar{k}_{opt}[j]$ first increases and then decreases to an asymptotic value after several bit intervals. This behavior is mainly due to the fact that, after several bit intervals, the ISI approaches an asymptotic value; see \cite{NoelJ3}. The slight difference between the amplification factor for fixed-gain AF relaying, $\bar{k}_{opt}$, and the asymptotic values of $\bar{k}_{opt}[j]$ is due to the fact that $\bar{k}_{opt}[j]$ is averaged over all bit intervals to obtain $\bar{k}_{opt}$.        

In what follows, the simulated error probability is averaged over $3 \times 10^{4}$ random realizations of the sequence $\ws{S}{1}{L}$, and the expected error probability was evaluated via (\ref{Eq. ErrorExpSeq}), after taking into account the modifications required for each protocol. Furthermore, for a fair comparison of the performance of the network for different bit intervals, we assume that the frequency of sampling, $t_{0}$, and the number of samples per bit interval, $M$, are both independent of $T$, i.e., for any $T$ the samples are taken at times $t_{m} = \{ 20, 40, 60,\ldots , 200 \}$ $\mu$s within the current bit interval.

\begin{figure}[!t]
	\centering
	\includegraphics[scale=0.27]{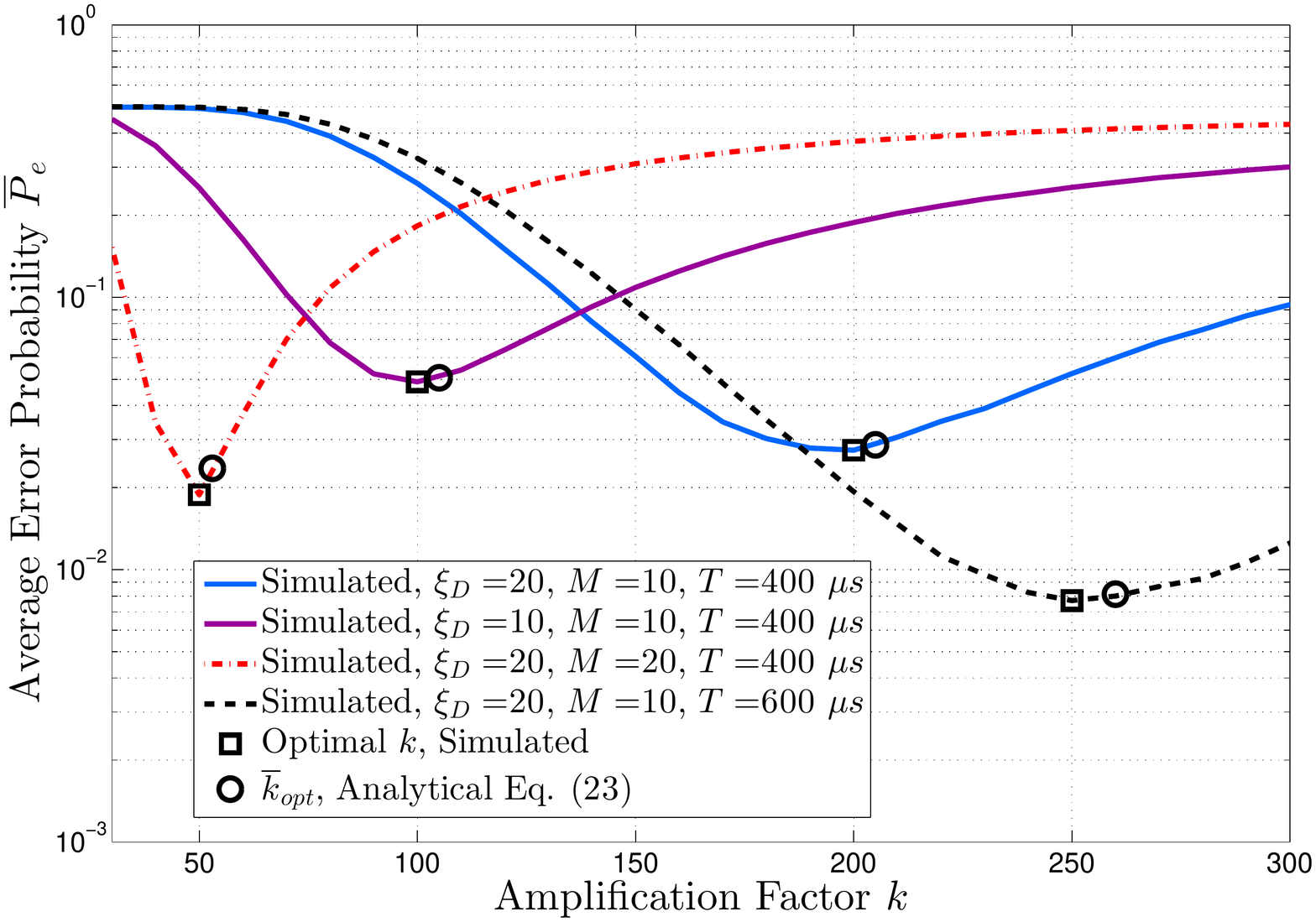}\vspace*{-6mm}
	\caption{Average error probability of a two-hop link as a function of the amplification factor $k$.}
	\label{Fig.4}
	\vspace*{-5 mm}
\end{figure}  
In Fig. \ref{Fig.4}, we evaluate the average error probability of a two-hop network with fixed-gain AF relaying to assess the accuracy of the optimal value $\bar{k}_{opt}$. We evaluated $\bar{P}_{e}$ for different system parameters, i.e., $M = \{10,20\}$, $T = \{400, 600 \} \, \mu s$, and $\xi_{D} = \{10, 20\}$. We observe that, by doubling $\xi_{D}$, and keeping the other parameters constant, the optimal $k$ is approximately doubled which is in agreement with (\ref{Eq. OptimalK}), when $P_{1}=P_{0}=0.5$. We can also see that increasing $M$ and $T$ decreases and increases the optimal $k$, respectively. Finally, we note the excellent match between the optimal $k$ observed via simulation and that derived analytically.    

\begin{figure}[!t]
	\centering
	\vspace*{-4mm}
	\includegraphics[scale=0.27]{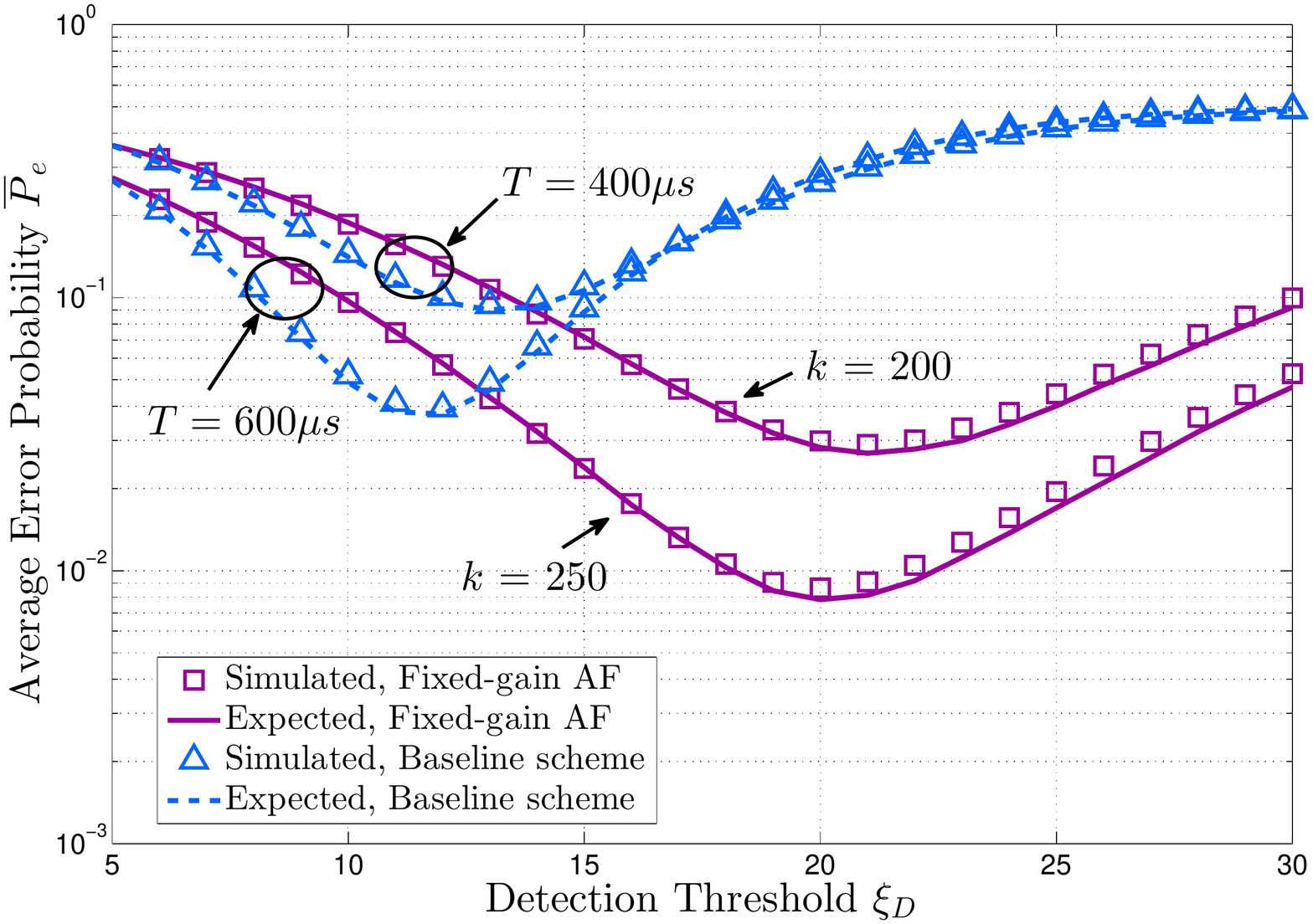}\vspace*{-6mm}
	\caption{Average error probability of fixed-gain AF relaying and the baseline scheme vs. the detection threshold $\xi_{D}$.}
	\label{Fig.5}
	\vspace*{-8 mm}
\end{figure}
Fig. \ref{Fig.5} shows the average error probability of fixed-gain AF relaying and the baseline case as a function of detection threshold $\xi_{D}$ for two different sets of system parameters, i.e., $ \{M=10, T=400 \, \mu s, k=200 \}$ and $ \{M=10, T=600 \, \mu s, k=250 \}$. The results show that fixed-gain AF relaying improves the overall performance of the network. We also see that increasing $T$ improves the performance of fixed-gain AF relaying, since increasing $T$ reduces the effect of ISI which, in turn, decreases the effect of amplified ISI at the destination node. Furthermore, a comparison of the results in Fig. \ref{Fig.4} and Fig. \ref{Fig.5} reveals that the optimization of the detection thresholds at node $D$ for a given $k$ is equivalent to optimizing $k$ for a given $\xi$.
        
\begin{figure}[!t]
	\centering
	\includegraphics[scale=0.27]{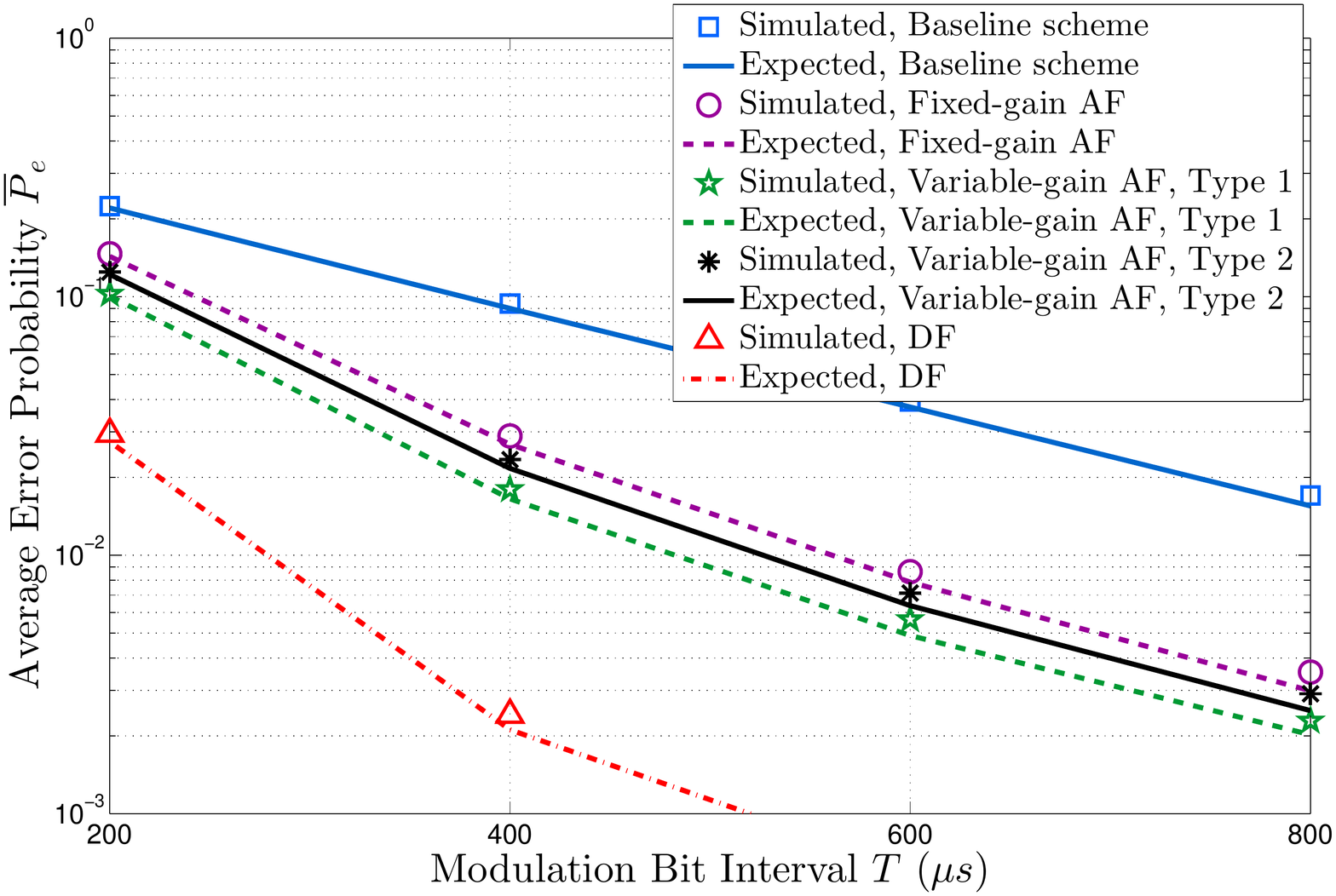}\vspace*{-6mm}
	\caption{Average error probability of fixed- and variable-gain AF relaying, DF relaying, and the baseline case as a function of modulation bit interval $T$.}
	\label{Fig.6}
	\vspace*{-5 mm}
\end{figure}  
In Fig. \ref{Fig.6}, the average error probability of a two-hop network for fixed- and variable-gain AF relaying, the DF relaying scheme from \cite{Arman2, Arman1}, and the baseline case are evaluated as a function of modulation bit interval $T$. $\bar{k}_{opt}$, $\hat{k}_{opt}[\cdot]$, and $\bar{k}_{opt}[\cdot]$ are applied for fixed-gain, Type 1 variable-gain, and Type 2 variable-gain relaying, respectively. The results show that all considered AF relaying protocols outperform the baseline scheme where the performance gap increases with $T$. Both types of variable-gain AF relaying perform slightly better than fixed-gain AF relaying, since the amplification factor is adjusted in each bit interval according to the current optimal amplification factor based on the expected ISI. Furthermore, we observe that DF relaying outperforms AF relaying. This is because in DF relaying, the ISI of the first hop is not amplified by the relay node. However, the improved performance of DF relaying comes at the expense of an increased complexity as the relay has to decode the message received from the source, which is not necessary for fixed-gain and Type 2 variable-gain AF relaying. Hence, these schemes are more suitable for nodes with limited processing capability. 
\section{CONCLUSION}
\label{Sec.Con} 
In this paper, we considered a three-node network where a nano-relay is deployed between a nano-transmitter and a nano-receiver. We assumed that the nano-transmitter emits \emph{multiple} \emph{random} bits, and proposed two relaying protocols, namely fixed- and variable-gain AF relaying. Furthermore, we derived closed-form expressions for the optimal amplification factors at the relay node for minimization of the expected error probability of the network. We showed via simulation and analysis that AF relaying improves the overall performance of the network. In particular, fixed-gain and Type 2 variable-gain AF relaying are attractive for implementation when the relay node has limited computational capabilities.    

\bibliography{IEEEabrv,Library}

\begin{thebibliography}{10}
\providecommand{\url}[1]{#1}
\csname url@samestyle\endcsname
\providecommand{\newblock}{\relax}
\providecommand{\bibinfo}[2]{#2}
\providecommand{\BIBentrySTDinterwordspacing}{\spaceskip=0pt\relax}
\providecommand{\BIBentryALTinterwordstretchfactor}{4}
\providecommand{\BIBentryALTinterwordspacing}{\spaceskip=\fontdimen2\font plus
\BIBentryALTinterwordstretchfactor\fontdimen3\font minus
  \fontdimen4\font\relax}
\providecommand{\BIBforeignlanguage}[2]{{%
\expandafter\ifx\csname l@#1\endcsname\relax
\typeout{** WARNING: IEEEtran.bst: No hyphenation pattern has been}%
\typeout{** loaded for the language `#1'. Using the pattern for}%
\typeout{** the default language instead.}%
\else
\language=\csname l@#1\endcsname
\fi
#2}}
\providecommand{\BIBdecl}{\relax}
\BIBdecl

\bibitem{NakanoB}
T.~Nakano, A.~W. Eckford, and T.~Haraguchi, \emph{Molecular
  Communication}.\hskip 1em plus 0.5em minus 0.4em\relax Cambridge University
  Press, 2013.

\bibitem{Raven}
P.~H. Raven and G.~B. Johnson, \emph{Biology}.\hskip 1em plus 0.5em minus
  0.4em\relax McGraw-Hill Science/Engineering/Math, 2001.

\bibitem{Einolghozati1}
A.~Einolghozati, M.~Sardari, A.~Beirami, and F.~Fekri, ``Data gathering in
  networks of bacteria colonies: Collective sensing and relaying using
  molecular communication,'' in \emph{Proc. IEEE INFOCOM}, Mar. 2012, pp.
  256--261.

\bibitem{Einolghozati2}
A.~Einolghozati, M.~Sardari, and F.~Fekri, ``Relaying in diffusion-based
  molecular communication,'' in \emph{Proc. IEEE ISIT}, Jul. 2013, pp.
  1844--1848.

\bibitem{Nakano2}
T.~Nakano and J.~Shuai, ``Repeater design and modeling for molecular
  communication networks,'' in \emph{Proc. IEEE INFOCOM}, Apr. 2011, pp.
  501--506.

\bibitem{Nakano3}
T.~Nakano and J.-Q. Liu, ``Design and analysis of molecular relay channels: An
  information theoretic approach,'' \emph{{IEEE} Trans. Nanobiosci.}, vol.~9,
  no.~3, pp. 213--221, Sep. 2010.

\bibitem{Unluturk1}
B.~Unluturk, D.~Malak, and O.~Akan, ``Rate-delay tradeoff with network coding
  in molecular nanonetworks,'' \emph{{IEEE} Trans. Nanotechnol.}, vol.~12,
  no.~2, pp. 120--128, Mar. 2013.

\bibitem{Balasubramaniam1}
S.~Balasubramaniam and P.~Lio, ``Multi-hop conjugation based bacteria
  nanonetworks,'' \emph{{IEEE} Trans. Nanobiosci.}, vol.~12, no.~1, pp. 47--59,
  Mar. 2013.

\bibitem{Walsh1}
F.~Walsh and S.~Balasubramaniam, ``Reliability and delay analysis of multihop
  virus-based nanonetworks,'' \emph{{IEEE} Trans. Nanotechnol.}, vol.~12,
  no.~5, pp. 674--684, Sep. 2013.

\bibitem{Bazargani1}
M.~Bazargani and D.~Arifler, ``Deterministic model for pulse amplification in
  diffusion-based molecular communication,'' \emph{{IEEE} Commun. Lett.},
  vol.~18, no.~11, pp. 1891--1894, Nov 2014.

\bibitem{Arman2}
\BIBentryALTinterwordspacing
A.~Ahmadzadeh, A.~Noel, and R.~Schober, ``Analysis and design of multi-hop
  diffusion-based molecular communication networks,'' \emph{Accepted subject to
  minor revisions, IEEE J. Sel. Areas Commun.}, 2014. [Online]. Available:
  \url{arXiv:1410.5585}
\BIBentrySTDinterwordspacing

\bibitem{Arman1}
------, ``Analysis and design of two-hop diffusion-based molecular
  communication networks,'' in \emph{Proc. IEEE GLOBECOM}, Dec 2014, pp.
  2820--2825.

\bibitem{NoelJ2}
A.~Noel, K.~C. Cheung, and R.~Schober, ``Optimal receiver design for diffusive
  molecular communication with flow and additive noise,'' \emph{{IEEE} Trans.
  Nanobiosci.}, vol.~13, no.~3, pp. 350--362, Sept 2014.

\bibitem{NoelJ1}
------, ``Improving receiver performance of diffusive molecular communication
  with enzymes,'' \emph{{IEEE} Trans. Nanobiosci.}, vol.~13, no.~1, pp. 31--43,
  Mar. 2014.

\bibitem{NoelPro1}
------, ``Using dimensional analysis to assess scalability and accuracy in
  molecular communication,'' in \emph{Proc. IEEE ICC MONACOM}, Jun. 2013, pp.
  818--823.

\bibitem{Ilienko}
A.~Ilienko, ``Continuous counterparts of poisson and binomial distributions and
  their properties,'' \emph{Annales Univ. Sci. Budapest, Sect. Comp.}, vol.~39,
  p. 137–147, 2013.

\bibitem{abramowitz}
M.~Abramowitz and I.~Stegun, \emph{Handbook of Mathematical Functions},
  1st~ed.\hskip 1em plus 0.5em minus 0.4em\relax New York: Dover, 1964.

\bibitem{NoelJ3}
A.~Noel, K.~C. Cheung, and R.~Schober, ``A unifying model for external noise
  sources and {ISI} in diffusive molecular communication,'' \emph{{IEEE} J.
  Sel. Areas Commun.}, vol.~32, no.~12, pp. 2330--2343, Dec 2014.

\end{thebibliography}

\end{document}